\documentclass[submission,copyright,creativecommons]{eptcs}
\providecommand{\event}{SOS 2007} % Name of the event you are submitting to

\usepackage{iftex}

\ifpdf
  \usepackage{underscore}         % Only needed if you use pdflatex.
  \usepackage[T1]{fontenc}        % Recommended with pdflatex
\else
  \usepackage{breakurl}           % Not needed if you use pdflatex only.
\fi

\usepackage{booktabs}
\usepackage{dsfont}
\usepackage{textcomp}
\usepackage[utf8]{inputenc} 
\usepackage[T1]{fontenc}
\usepackage{xspace}
\usepackage{longtable}
\usepackage[table,xcdraw]{xcolor}
\usepackage{graphicx} % Required for inserting images
\usepackage{iftex}
\usepackage{listings}
\usepackage{xcolor}
\usepackage{makecell}
\usepackage{amsmath}
\usepackage{graphicx}
\usepackage{multirow}
\usepackage{multicol}
\definecolor{linkcolor}{HTML}{A93C93}
\PassOptionsToPackage{obeyspaces}{url}%
\usepackage{url}

\usepackage[
textwidth=0.125\textwidth,textsize=footnotesize]{todonotes}
\usepackage{relsize}
\newcommand{\code}{\lstinline[style=MyInline]}
\definecolor{PrologPredicate}{RGB}{0,0,200}
\definecolor{PrologVar}      {RGB}{145,032,039}
\definecolor{PrologComment}  {RGB}{169,082,044}
\lstdefinestyle{MyInline}
{
  basicstyle = \relsize{-0.5}\ttfamily\color{PrologPredicate},
  language=Prolog,
  mathescape = true,  
  breaklines = true,
  breakatwhitespace=true,
  upquote = true,
  literate =
  {?-}{{?-\,}}3
  {:-}{{:-\,}}2
  {.=.}{{\,\#=\,}}3
  {.<.}{{\,\#<\,}}3
  {.>.}{{\,\#>\,}}3
  {.=<.}{{\,\#=<\,}}4
  {.>=.}{{\,\#>=\,}}4
}
\lstdefinestyle{prolog}
{
    xleftmargin=0.5cm,
    numberstyle=\tiny\color{gray},
    numbers=left,
    stepnumber=1,
    language=Prolog,
    basewidth = 0.50em,
    basicstyle = \relsize{-0.5}\ttfamily\color{PrologPredicate},
    moredelim = {*[s][\color{PrologVar}]{(}{)}},
    commentstyle = \mdseries\color{PrologComment},
    morecomment=[l]\%,    
    literate =
        {:-}{{\textcolor{black}{:-}}}2
        {,}{{\textcolor{black}{,}}}1
        {.}{{\textcolor{black}{.}}}1,
}
\newcommand{\myurl}[1]{\href{https://github.com/Alexandara/vecsr/blob/main/#1}{\nolinkurl{#1}}}

\usepackage{fontawesome}
\newcommand{\breakfastlinkv}{\href{https://github.com/Alexandara/vecsr/blob/main/Examples/Breakfast/breakfast_vecsr_solution.txt}{\faExternalLink~}}
\newcommand{\breakfastlinkg}{\href{https://github.com/Alexandara/vecsr/blob/main/Examples/Breakfast/breakfast_gpt_solution.txt}{\faExternalLink~}}
\newcommand{\breakfastlinkf}{\href{https://github.com/Alexandara/vecsr/blob/main/Examples/Breakfast/}{\faExternalLinkSquare~}}

\newcommand{\gotosleeplinkv}{\href{https://github.com/Alexandara/vecsr/blob/main/Examples/Go_To_Sleep/go_to_sleep_vecsr_solution.txt}{\faExternalLink~}}
\newcommand{\gotosleeplinkg}{\href{https://github.com/Alexandara/vecsr/blob/main/Examples/Go_To_Sleep/go_to_sleep_gpt_solution.txt}{\faExternalLink~}}
\newcommand{\gotosleeplinkf}{\href{https://github.com/Alexandara/vecsr/blob/main/Examples/Go_To_Sleep/}{\faExternalLinkSquare~}}

\newcommand{\browseinternetlinkv}{\href{https://github.com/Alexandara/vecsr/blob/main/Examples/Browse_Internet/browse_internet_vecsr_solution.txt}{\faExternalLink~}}
\newcommand{\browseinternetlinkg}{\href{https://github.com/Alexandara/vecsr/blob/main/Examples/Browse_Internet/browse_internet_gpt_solution.txt}{\faExternalLink~}}
\newcommand{\browseinternetlinkf}{\href{https://github.com/Alexandara/vecsr/blob/main/Examples/Browse_Internet/}{\faExternalLinkSquare~}}

\newcommand{\washteethlinkv}{\href{https://github.com/Alexandara/vecsr/blob/main/Examples/Wash_Teeth/wash_teeth_vecsr_solution.txt}{\faExternalLink~}}
\newcommand{\washteethlinkg}{\href{https://github.com/Alexandara/vecsr/blob/main/Examples/Wash_Teeth/wash_teeth_gpt_solution.txt}{\faExternalLink~}}
\newcommand{\washteethlinkf}{\href{https://github.com/Alexandara/vecsr/blob/main/Examples/Wash_Teeth/}{\faExternalLinkSquare~}}

\newcommand{\brushteethlinkv}{\href{https://github.com/Alexandara/vecsr/blob/main/Examples/Brush_Teeth/brush_teeth_vecsr_solution.txt}{\faExternalLink~}}
\newcommand{\brushteethlinkg}{\href{https://github.com/Alexandara/vecsr/blob/main/Examples/Brush_Teeth/brush_teeth_gpt_solution.txt}{\faExternalLink~}}
\newcommand{\brushteethlinkf}{\href{https://github.com/Alexandara/vecsr/blob/main/Examples/Brush_Teeth/}{\faExternalLinkSquare~}}

\newcommand{\vacuumlinkv}{\href{https://github.com/Alexandara/vecsr/blob/main/Examples/Vacuum/vacuum_vecsr_solution.txt}{\faExternalLink~}}
\newcommand{\vacuumlinkg}{\href{https://github.com/Alexandara/vecsr/blob/main/Examples/Vacuum/vacuum_gpt_solution.txt}{\faExternalLink~}}
\newcommand{\vacuumlinkf}{\href{https://github.com/Alexandara/vecsr/blob/main/Examples/Vacuum/}{\faExternalLinkSquare~}}

\newcommand{\changesheetslinkv}{\href{https://github.com/Alexandara/vecsr/blob/main/Examples/Change_Sheets_and_Pillow_Cases/change_sheets_and_pillow_cases_vecsr_solution.txt}{\faExternalLink~}}
\newcommand{\changesheetslinkg}{\href{https://github.com/Alexandara/vecsr/blob/main/Examples/Change_Sheets_and_Pillow_Cases/change_sheets_and_pillow_cases_gpt_solution.txt}{\faExternalLink~}}
\newcommand{\changesheetslinkf}{\href{https://github.com/Alexandara/vecsr/blob/main/Examples/Change_Sheets_and_Pillow_Cases/}{\faExternalLinkSquare~}}

\newcommand{\washdirtydisheslinkv}{\href{https://github.com/Alexandara/vecsr/blob/main/Examples/Wash_Dirty_Dishes/wash_dirty_dishes_vecsr_solution.txt}{\faExternalLink~}}
\newcommand{\washdirtydisheslinkg}{\href{https://github.com/Alexandara/vecsr/blob/main/Examples/Wash_Dirty_Dishes/wash_dirty_dishes_gpt_solution.txt}{\faExternalLink~}}
\newcommand{\washdirtydisheslinkf}{\href{https://github.com/Alexandara/vecsr/blob/main/Examples/Wash_Dirty_Dishes/}{\faExternalLinkSquare~}}

\newcommand{\feedmelinkv}{\href{https://github.com/Alexandara/vecsr/blob/main/Examples/Feed_Me/feed_me_vecsr_solution.txt}{\faExternalLink~}}
\newcommand{\feedmelinkg}{\href{https://github.com/Alexandara/vecsr/blob/main/Examples/Feed_Me/feed_me_gpt_solution.txt}{\faExternalLink~}}
\newcommand{\feedmelinkf}{\href{https://github.com/Alexandara/vecsr/blob/main/Examples/Feed_Me/}{\faExternalLinkSquare~}}

\newcommand{\readlinkv}{\href{https://github.com/Alexandara/vecsr/blob/main/Examples/Read/read_vecsr_solution.txt}{\faExternalLink~}}
\newcommand{\readlinkg}{\href{https://github.com/Alexandara/vecsr/blob/main/Examples/Read/read_gpt_solution.txt}{\faExternalLink~}}
\newcommand{\readlinkf}{\href{https://github.com/Alexandara/vecsr/blob/main/Examples/Read/}{\faExternalLinkSquare~}}

\title{VECSR: Virtually Embodied \\Common Sense Reasoning System}
\author{Alexis R. Tudor
\institute{University of Texas at Dallas, USA}
\email{alexisrenee1@gmail.com}
\and
Joaqu\'in Arias
\institute{CETINIA \\ Universidad Rey Juan Carlos, Spain}
\email{joaquin.arias@urjc.es}
\and
Gopal Gupta
\institute{University of Texas at Dallas, USA}
\email{gupta@utdallas.edu}
}
\def\titlerunning{VECSR}
\def\authorrunning{A.R. Tudor, J. Arias \& G. Gupta}
\begin{document}
\maketitle

\begin{abstract}
  % Context
  The development of autonomous agents has seen a revival of
  enthusiasm due to the emergence of LLMs, such as GPT-4o.
  % Problem to be solved
  Deploying these agents in environments where they coexist with
  humans (e.g., as domestic assistants) requires special attention to
  trustworthiness and explainability.
  % SOTA & limitations
  However, the use of LLMs and other deep learning models still
  does not resolve these key issues. Deep learning systems may hallucinate, be unable to justify their decisions as black boxes, or perform badly on unseen scenarios.
  % Proposal
  In this work, we propose the use of s(CASP), a goal-directed common
  sense reasoner based on Answer Set Programming, to break down the
  high-level tasks of an autonomous agent into mid-level instructions
  while justifying the selection of these instructions.
  % Evaluation (validation)
  To validate its use in real applications we present a
  framework that integrates the reasoner into the VirtualHome
  simulator and compares its accuracy with GPT-4o, running some of the
  ``real'' use cases available in the domestic environments of VirtualHome.
  Additionally, since experiments with VirtualHome have shown the need
  to reduce the response time (which increases as the agent's decision
  space grows), we have proposed and evaluated a series of
  optimizations based on program analysis that exploit the advantages
  of the top-down execution of s(CASP).
\end{abstract}

\section{Introduction}

As autonomous aids, both robotic and digital, have become more commonplace, a need has surfaced for more trustworthy and generalizable task breakdown. Tasks which humans can perform instinctively (e.g., make a sandwich) need to be broken down into composite executable mid-level steps by an autonomous agent (grab bread, grab peanut butter, etc.). To accomplish this breakdown, modern research uses deep learning systems \cite{morales2021} that consume large quantities of data to replicate actions and patterns represented in that data. Large language models (LLMs) in particular have grown in both function and popularity in recent years. However, LLMs and other deep learning-based systems do not reason the same way that humans do and thus are prone to hallucinations and mistakes unlike those that would be made by a human. Even when it is well known that a certain model will produce erroneous output, the models often cannot explain their decisions in a way that allows for correction, leading to new vulnerabilities \cite{spracklen2025}. This lack of explainability leads to lower trust in the model's decisions \cite{gunning2021}.

Furthermore, a recent study by the Association for the Advancement of Artificial Intelligence \cite{aaai2025} found that 76\% of AI professionals surveyed believed that current approaches were ``unlikely'' or ``very unlikely'' to yield true artificial general intelligence. Despite these notable downsides in the use of deep learning systems, LLMs have continued to proliferate. Beyond the generation of natural language, researchers are searching for ways to use LLMs in embodied environments. That is an environment, simulated or real, where an autonomous agent is ``embodied'', that is, able to move around the space and interact with it. Research into using LLMs for the control of robotics is already underway, but continues to struggle with the aforementioned problems as noted in a survey by Wang et al. \cite{wang2025}. Such systems have already been explored in virtual environments as in the research by Huang et al. \cite{huang2022} with mixed results. We posit that for trustworthy autonomous models, a new approach is required. We propose a symbolic and explainable approach to common sense reasoning using s(CASP), a goal-directed constraint answer set programming system that allows for powerful and justifiable reasoning.

We demonstrate the use of s(CASP) for common sense reasoning on the problem of breaking down high-level tasks into mid-level instructions and compare it to results using LLMs. We will first provide background on relevant prior works in Section \ref{sec:background}, including the simulated environment used for validation and the metrics proposed by Huang et al. \cite{huang2022} that we will be using for comparison to the LLM system for task completion. In that section we also discuss s(CASP) and a brief comparison to other planning solutions. Our system, which we call Virtually Embodied Common Sense Reasoning (VECSR), is described in Section \ref{sec:method}. To use a symbolic system for reasoning in an embodied environment is a novel approach. We describe our method of modeling real-world information in s(CASP) in Section \ref{sec:template}, breaking that information into actionable tasks in Section \ref{sec:generation}, and our approach for overcoming the unique challenges associated with the large state in Section \ref{sec:static}. 

We compare the results of VECSR to the research of Huang et al. \cite{huang2022} and to the more modern GPT-4o LLM in Section \ref{sec:results}. We use the metrics of correctness and executability proposed by Huang et al. \cite{huang2022}, and evaluate the system for generalizability to unseen data and time of execution. Our system creates action plans that are more accurate and executable than those created by LLMs while also providing benefits entirely absent in LLM results, such as context-awareness and justifiability.

There are three main contributions of this paper. Firstly, the creation of VECSR,  a novel framework that accepts high-level tasks and uses s(CASP) to plan and execute those tasks in a context-aware fashion in a fully-embodied simulated environment. Secondly, optimization tools based on statically analyzing the program at compile time to improve the execution time and tractability of s(CASP) in real-world environments. And lastly, an evaluation of VECSR's correctness and executability versus GPT-4o, an analysis of VECSR's generalizability, and validation of the optimizations in terms of time of execution.

\section{Background and Related Works}
\label{sec:background}

\subsection{VirtualHome: The Simulation Environment}
\label{sec:sim}
\begin{figure*}[h]
\centering
\includegraphics[width = 300px]{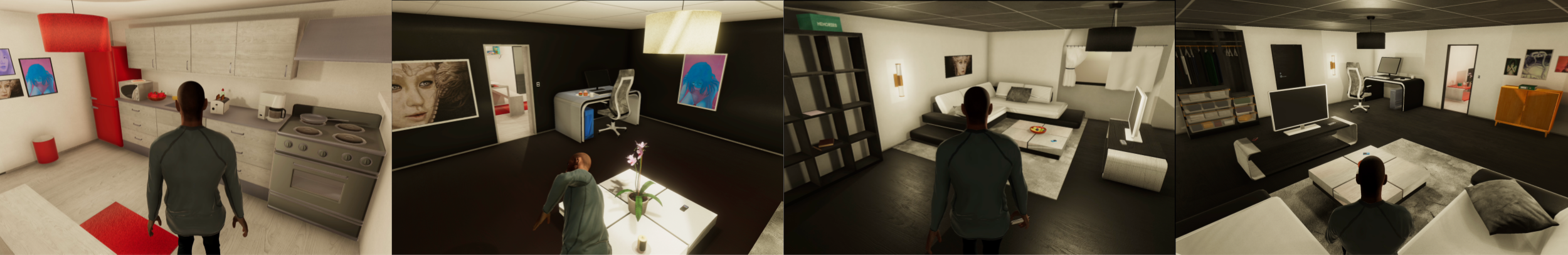}
\caption{The VirtualHome simulated environment provides a variety of apartment layouts for embodied agents to interact with.}
\label{fig:vh}
\end{figure*}

For demonstration of VECSR we use the VirtualHome simulated environment shown in Figure \ref{fig:vh}. VirtualHome provides several simulated domestic environments with an API through which one or more agents may take actions and retrieve the state (available at \url{http://virtual-home.org}). It comes with common sense knowledge about each object (whether it is grabbable, movable, edible, etc.), information about the current state of the object (turned on, opened or closed, etc.), and its relationship with other objects (on top of, close to, inside of, etc.). The primary advantage of VirtualHome over other simulation engines is that it uses a mid-level control schema. This means that we can give the agent commands like ``grab remote'' rather than dealing with the details of actual movement that would be more appropriate for a robotic controller. For our research, focused on the common sense reasoning portion of the task-completion process, mid-level control allows for reasoning at the level of the human mind. 

The creators of VirtualHome have provided a large human-created dataset of decomposed tasks in the format expected by the simulator. The dataset contains 549 activities, each with one or more human-created breakdowns into mid-level executable units that lead to the completion of those tasks. It is an excellent source of tasks to test various common sense reasoning systems proposed. Note that there are more objects and actions present in the dataset than are in the simulation itself. However, VirtualHome remains an excellent tool for demonstration of common sense reasoning in an embodied environment. We provide several video examples of tasks completed in the VirtualHome environment at \url{https://drive.google.com/drive/folders/154-KlMgywB1arc9xFQLpAs1K2i9t51j8?usp=sharing}.

\subsection{Metrics to Evaluate Task Completion}
\label{sec:metrics}

Given a task, our goal is to find a sequence of actions that act as instructions to the autonomous agent to complete that task. Huang et al. \cite{huang2022} have proposed two metrics for evaluating this sequence of actions: correctness and executability. Correctness is the measurement of whether a sequence of actions accomplishes the task. To evaluate this metric, Huang et al. had human participants review the generated action sequences and assess whether they believed the sequence of actions was correct or not. Additionally, they speculate that correctness can be measured by comparing the final state reached by the action sequence to that reached by a human completing that same task. Huang et al. stated that this may not be possible due to a need to keep the initial state constant between tasks. This is a limitation of the LLM, which cannot accept the initial state data except as part of the prompt, which leads to unreliable reproducability of state. Our methodology is context-aware, ensuring a repeatable initial state. This second method for evaluating correctness is used by VECSR when creating plans.

As defined by Huang et al., a set of instructions is executable if it can be correctly parsed and it satisfies the common sense constraints of the system. For parsing, each instruction must meet the syntactical requirements of the VirtualHome system. Satisfying the common sense constraints of the system means that all pre-conditions of each action are met prior to taking the action. The example used highlights that the agent cannot grab milk out of the fridge if the fridge door is still shut. In addition to physical constraints specified by the VirtualHome system, there are also non-physical constraints. For instance, a food like chicken cannot be safely eaten without being cooked. These secondary constraints are easily ignored by the LLM system because of lack of context-awareness. Therefore, instead of being aware that the only food that is in the house is chicken, it would propose an instruction set that has the agent put ``food'' on a plate and then eat the ``food''. Even when such valuable context is provided in the prompt to the LLM, it is still prone to hallucination of additional state information or forgetting existing information over multiple instructions.

In Section \ref{sec:gpt4o} we use these metrics, correctness and executability, to compare VECSR to an equivalent LLM reasoner.

\subsection{s(CASP) vs.\ Other Planning Systems}
\label{sec:scasp}
Logic programming approaches have several key advantages over deep learning systems. They are inherently explainable, can produce  comparable results (see Gupta et al. \cite{gupta2023}) for a specific domain, and are compact and logically sound. We use s(CASP) by Arias et al. \cite{arias2018}, a goal-directed constraint-based answer set programming (ASP) system that executes ASP programs in a top-down manner similar to Prolog, eliminating the need for grounding. This allows s(CASP) to reason over incomplete information. s(CASP) outputs partial answer sets containing only the information needed to successfully execute a query.

Because of the large number of facts and rules associated with even simple real-world environments, being able to create partial answer sets is important from an efficiency viewpoint. Additionally, s(CASP) allows for human-like reasoning with classical negation, negation-as-failure, default rules, and even and odd loops over negation. These features become critical for practical task planning, such as in the example below. Consider the following facts:
\begin{lstlisting}
grabbable(mug).
movable(mug).
grabbable(bowl).
\end{lstlisting}
Expressing the common sense knowledge that normally things that are grabbable are also movable is difficult to express in classical logic, but easily expressed in ASP and s(CASP). This is accomplished by rules such as \code{movable(X) :- grabbable(X), not -movable(X)}, which states that \code{movable(bowl)} holds if \code{grabbable(bowl)} holds and \code{-movable(bowl)} is not explicitly asserted. 

We view the problem of breaking down a task into a sequence of actions as a planning problem. This approach is similar to other planning-based solutions such as those using PDDL \cite{ghallab1998} or SOAR \cite{laird2019}. The s(CASP) template used is similar in structure to a PDDL program, with the initial state (derived from the simulated environment, see Section \ref{sec:sim}), a set of possible actions along with preconditions and effects of those actions, and a goal defined in terms of a desired end state. However, reasoning in PDDL is constrained as it only supports conservative reasoning without abducibles (used in the example above) and even loops. ASP provides significantly more flexibility, especially for possible future research, such as inferring generalized knowledge from a specific task.

The SOAR Cognitive Architecture comes with its own programming language, however like other works we seek to leverage its architecture without using the SOAR language \cite{sumers2024}. Our architecture uses a similar model as SOAR of keeping a working memory loaded while only selecting relevant items from long-term memory as needed. The use of s(CASP) as a reasoner in our case allows for the strengths of SOAR's design combined with the powerful reasoning mentioned in Section \ref{sec:scasp}. 

\section{Design of VECSR: Virtually Embodied Common Sense Reasoning}
\label{sec:method}

\begin{figure*}
\centering
\includegraphics[width = 300px]{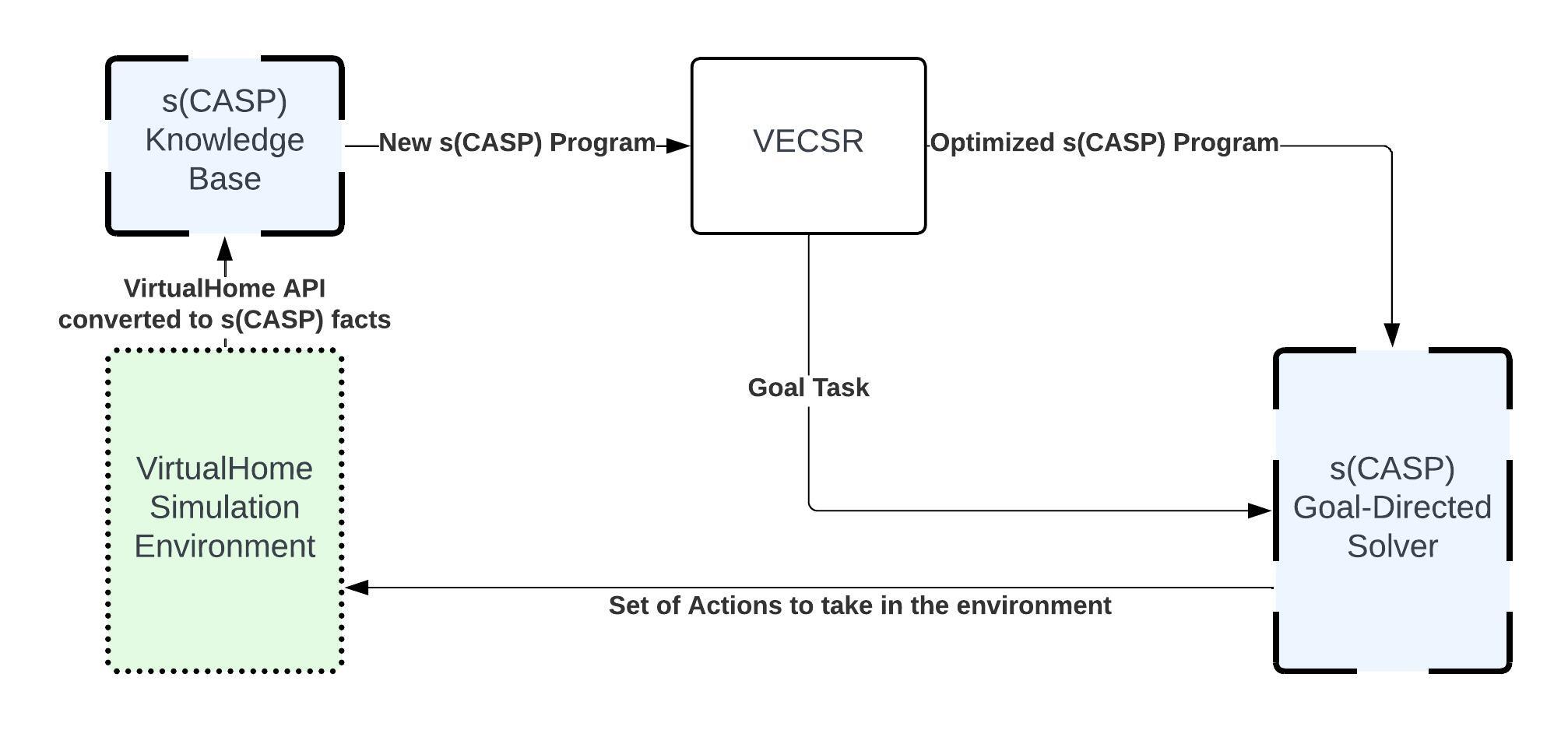}
\caption{VECSR: framework for s(CASP) common sense reasoning in the VirtualHome simulator.}
\label{fig:scaspharness}
\end{figure*}

The design for VECSR, including the connection to the VirtualHome simulation environment, is illustrated in Figure \ref{fig:scaspharness}. The resulting flowchart consists of five main stages:
\begin{enumerate}
    \item Converting the VirtualHome state to s(CASP) facts.
    \item Combining the state facts with common sense rules and constraints.
    \item Optimizing the newly created program.
    \item Solving an appropriate query to create an action sequence for a goal task.
    \item Executing the instructions.
\end{enumerate}

After step five, the cycle can start again. This section describes the way we convert the VirtualHome state into s(CASP) facts (Section \ref{sec:template}), the compile-time optimization performed on the knowledge base (Section \ref{sec:static}), and how we generate the action plans (Section \ref{sec:generation}). The full code for VECSR is available at \url{https://github.com/Alexandara/vecsr}.

\subsection{Modeling VirtualHome Environments in s(CASP)}
\label{sec:template}

A notable weakness of symbolic methods is the representation of large amounts of real-world data for reasoning by a symbolic solver. We have undergone several iterations of s(CASP) templates to represent the simulation environment in a way that is conducive for fast and accurate processing. The VirtualHome API provides data on the environment in the form of a knowledge graph. In this knowledge graph, each object is a node and relationships between objects are edges. In each node there is a variety of state information, which we break into four classes: type, attribute, state, and category. As an example, a specific lamp (lamp1) can have the attribute of having a switch, its state could be off or on, and it could belong to the category of light sources. That information in s(CASP) is:
\begin{lstlisting}
type(lamp1, lamp).
has_switch(lamp1).
on(lamp1).
lightsource(lamp1).
\end{lstlisting}

We make the simplifying assumption that type, attribute, and category will not change. However, the state of an object could change in the course of an action plan being carried out (in this example, the lamp could be switched off). The same problem occurs with relationships between objects. If the lamp is on a table, \code{on_top_of(lamp1, table2)}, it can be removed, changing the state. To prevent the old state from clogging up the program during the dynamic part of execution, we use a list to store changeable state:
\begin{lstlisting}
on([lamp1, lamp3]).
on_top_of([[lamp1, table2], [lamp3, floor4]]).
\end{lstlisting}

This allows the program to maintain the state dynamically while calculating prospective action plans that may accomplish the goal. These facts are then combined with the common sense knowledge base and planning problem rules defined in the next section. 

\subsection{Breaking Down Tasks into Instructions}
\label{sec:generation}

We approach the problem of action plan generation as a traditional planning problem. The foundation of this method is that the program starts with an initial state and a final state and attempts to find a plan that satisfies transforming the initial state into the final state. The main loop is as follows:
\begin{lstlisting}
transform(FinalState, Plan) :- 
    initial_state(State1), transform(State1, FinalState, [State1], Plan).
transform(State1, FinalState, _, []) :- 
    state_subset(FinalState, State1).
transform(State1, FinalState, Visited, [Action|Actions]) :-
    choose_action(Action, State1, FinalState),
    update(Action, State1, State),
    not_member(State, Visited),
    transform(State, FinalState, [State|Visited], Actions).
\end{lstlisting}

Because the state is described as predicates that are computed dynamically, the initial state is converted into the format expected by the main loop. The initial state rule is in the following format \code{[ holds([$\dots$]), close([$\dots$]), on_top_of([$\dots$]) | _ ]} where each factor of interest has a list of items that match those parameters. Note that we use lists, \code{[A, B, C | _]}, where the number of items is defined at run time according to the simulated environment. An example where the agent is carrying a knife and has set a carrot on a cutting board would look like \code{[ holds([knife1]),} \code{close([knife1, carrot2, cuttingboard3, counter4]),} \code{on_top_of([[carrot2,cuttingboard3], [cuttingboard3,counter4]]) | _ ]}.\footnote{In this paper we will only display the relevant subset of the state for the example, not the full state as would be kept in the s(CASP) knowledge base.} Unlike in a traditional planning problem where the actions must lead to the exact final state requested, with VECSR the final state simply needs to be a subset of the state reached by the action plan. For example, if the final state is ``holding a phone'' then any final state where that is true will be accepted, whether the person is sitting or standing in the final state does not matter.

The actions are selected based on suggestions for the most optimal action to take for any given current and the final state and narrowed down by whether that action is legal or not. If the final state contains the agent sitting on the couch, then the action that completes that goal is to sit on the couch. But the action is not legal unless the agent is standing close to the couch. Legality is defined both by the preconditions of any given action and by the constraints in the knowledge base. So the action that is closest to the final action while still being legal is chosen at every step until the task is complete. The tasks themselves are defined by the final state of the world after the task is complete, meaning that every finite action plan is correct by definition. 

We used ten example tasks from the VirtualHome dataset to create the s(CASP) knowledge base. These ten example tasks were chosen based on the first ten example tasks in the paper by Huang et al. \cite{huang2022}.  
The small number of the sample was due to two factors: the amount of time it took to create a common sense knowledge base for a task and the incredible generalization ability of the s(CASP) program (discussed more in Section \ref{sec:generalizability}). The planning problem model augmented by dynamic simulated data makes a robust method for action generation based on compact common sense knowledge and reasoning with s(CASP). A small example representing the method for generating action plans is available\footnote{This file and the rest of examples are linked to \url{https://github.com/Alexandara/vecsr}.} at \myurl{Examples/example_plan.pl} in the repository.

\subsection{Optimization With Static Analysis Techniques}
\label{sec:static}

When first converting the simulation to s(CASP) facts, we had nearly 3,000 lines of facts by themselves. This was impractical and computationally complex as even the simplest of tasks took over three minutes to accomplish and the more complicated tasks took at least multiple days. For humans, reasoning happens in a matter of seconds, not minutes or hours. To improve the computation time of our system, we used three different compile-time optimizations, exploiting the query driven strategy of s(CASP), to prune the knowledge base. The smaller and more relevant the knowledge base, the quicker the processing time.

\textbf{Modular:} The first optimization we implemented was the use of proximity-based ``modules'' where the s(CASP) knowledge base contains only the state information from relevant areas of the simulation. This kind of analysis works on any data with the concept of ``proximity'' (such as knowledge graphs or tabulated data) but it works particularly well in an embodied environment where proximity is directly related to physical distance. We segmented the VirtualHome apartment into modules of facts based on rooms such that if the agent was completing a task such as ``Brush Teeth'' the s(CASP) knowledge base would only include state information for the bathroom. 

\textbf{Dependency Graph:} Secondly, to further reduce the knowledge base size, we create a dependency graph for the given query using the knowledge base rules, similar to that proposed by Nguyen et al. \cite{nguyen2008}. We create this dynamic dependency graph by starting from the  query goal as the root node. If goal G$_1$ calls goal G$_2$, then a dependency exists between G$_1$ and G$_2$. 
Rules that are untouched during the construction of a dependency graph are not considered during execution of the query. 

\textbf{Partial Grounding:} Finally, the most critical optimization is partial grounding of the program. Often, a task is ambiguous enough to be satisfied by multiple objects. If the task includes sitting, where? If the task includes reading, which book? The process of finding what entities the agent is going to use to fulfill a task and creating an action plan to accomplish the task can be separated from each other. First, the program locates the objects necessary for the task. Next, it eliminates any facts that do not pertain to those objects. As an example, if the task is to read then the agent would first select something in the simulation environment to read and then discard the rest. This optimization is similar to the modular optimization, but applied at a lower level taking the semantics of a primitive action (e.g., read) into account. Unlike the grounding done by ASP programs like clingo \cite{gebser2019}, this optimization uses constraints of the task to bind variables to objects in the state at compile-time, then only pulls the state regarding those objects from VirtualHome. 

\textbf{Fully Optimized:} Using these three optimizations together we are able to practically reason over the real-world state in near-real-time. Nevertheless, it is important to note that query-based execution of s(CASP) is what makes it feasible to apply models based on Answer Set Programming to the generation of plans in real-life scenarios.

% Example programs with each of the optimizations are available at \url{https://github.com/Alexandara/vecsr/} under the folder \myurl{Examples} where each sample task has its own folder. In each folder (which are named after the tasks) will be the following:
% \begin{itemize}
%     \item[] \code{task_standard.pl}: the original program for \code{task}.
%     \item[] \code{task_modular.pl}: the \code{task} program using ``Modular''.
%     \item[] \code{task_depgraph.pl}: the \code{task} program using ``Dependency Graph''.
%     \item[] \code{task_partground.pl}: the \code{task} program using ``Partial Grounding''.
%     \item[] \code{task_opt.pl}: the fully optimized \code{task} program.
% \end{itemize}

% These programs were dynamically generated using VECSR in combination with the VirtualHome simulation with instructions for generation available in \myurl{README.md} at the repository.

\section{Evaluation and Validation}
\label{sec:results}

To validate VECSR executability and correctness, we have conducted three evaluations: (i) comparison of the action plans generated with those generated by  GPT-4o (Section~\ref{sec:gpt4o}), (ii) correctness of VECSR's plans on unseen data/use cases (Section~\ref{sec:generalizability}), and (iii) run-time comparisons to measure the effect of our compile-time optimizations (Section~\ref{sec:time}). For the first and third evaluation, we use ten tasks listed by Huang et al. \cite{huang2022}. These range in complexity from ``Go To Sleep'', a simple task where the correct answer is to go lay in the bed, to ``Change Sheets in Pillow Case'' which requires multiple steps of removing one set of bedding and replacing it with another set. All ten examples are available at \url{https://github.com/Alexandara/vecsr/tree/main/Examples}, including the two aforementioned and the following: Browse Internet, Wash Teeth, Brush Teeth, Vacuum, Wash Dirty Dishes, Feed Me, Breakfast, and Read.

\subsection{Comparison to GPT-4o}
\label{sec:gpt4o}

LLMs are advancing rapidly, with improved LLMs being revealed multiple times a year (see a recent survey by Patil and Gudivada \cite{patil2024}). Since the paper by Huang et al. \cite{huang2022}, a new version of GPT was unveiled that performs better than GPT-3.5 on most standard benchmarks. We ran a similar user study as the one performed by Huang et al. on the newer GPT-4o for a fairer comparison using  correctness and execution-time as metrics. However, we do not evaluate the GPT-4o generated plans in terms of executability. The main contribution of the paper by Huang et al. was the use of a secondary LLM translation to create more executable plans, something we did not replicate. Without that additional translation, the executability of LLM-generated plans is very poor and would not make a good comparison metric. The highest executability the modified LLM system by Huang et al. was able to achieve was 78.57\%, and that executability rating ignored context-aware executability. Additionally, even on the best performing model, the LLM only made executable and correct plans 35.23\% of the time. VECSR plans are 100\% executable in real simulated environments and by definition are 100\% correct. VECSR tasks are defined by their post-conditions, and action plans are only accepted when the post-conditions of the task are complete. Combined with the strict checking of pre-condition-based legality of every action, this guarantees the final action plans are 100\% correct. 

\begin{figure*}
\centering
\includegraphics[width = 325px]{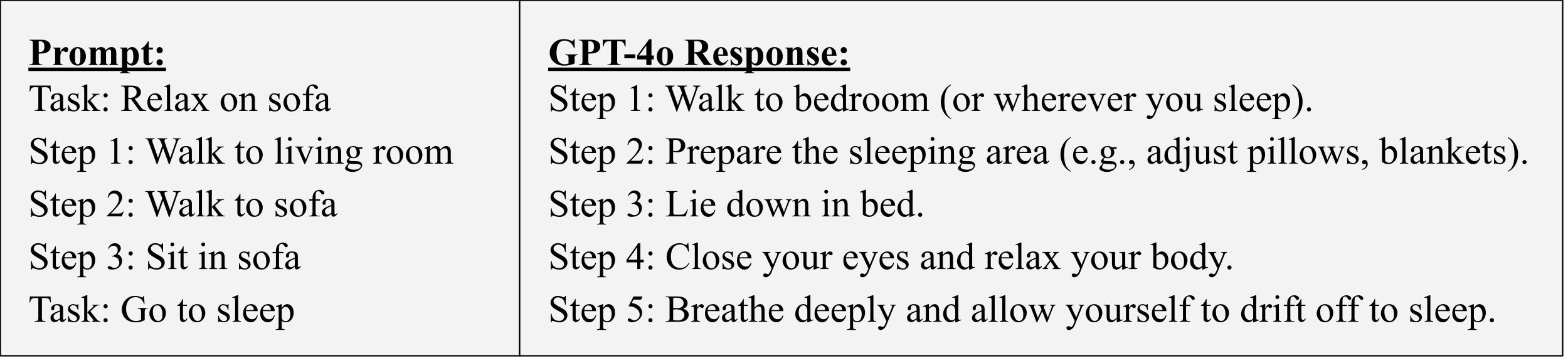}
\caption{Prompt where the task ``relax on sofa'' was used as an example for the task ``go to sleep''.}
\label{fig:gptprompt}
\end{figure*}

To use GPT-4o to generate action plans for tasks, we performed the same kind of zero-shot prompting as Huang et al. where we selected similar tasks from the database, like using the task ``relax on sofa'' as similar to the task ``go to sleep'', and prompted the model as in Figure \ref{fig:gptprompt}. The average length of the GPT-4o generated instructions was 12.3 steps compared to the 9.9 steps in the s(CASP) generated instructions. Huang et al. found that executable instruction sets were often shorter than unexecutable ones, and so the instruction sizes line up with the previous research. To evaluate the correctness of the GPT-4o generated action plans we performed a user study where 12 participants reviewed each of the ten tasks and marked the plans as correct or incorrect. Our results showed that GPT-4o performed similarly or better than the tested LLMs in the paper by Huang et al. and created correct plans in 65.91\% of cases. This is lower than the best GPT model used by Huang et al., however this is likely attributed to minor differences in the zero-shot prompting. LLMs are very susceptible to being influenced by the way prompts are done. As an experiment, we asked GPT-4o to provide instructions for the task ``go to sleep'' without providing the example zero-shot prompt, and received a task plan that was 20 steps and over 300 words in length.  Nonetheless, these results indicate that the results found by Huang et al. are still relevant for newer versions of GPT. Table \ref{tab:correct} shows more detailed information on the correctness and plan length of each task. Additionally, in the table each task has a link to the solutions generated by each system (the icons to the left of the tasks, first is the VECSR solution and second is the GPT-4o solution), which can be found in the \myurl{Examples} folder alongside the programs mentioned in Section \ref{sec:static}. These examples can also be dynamically generated using the VECSR system. 

\begin{table}
 \centering
 \caption{Number of steps generated (VECSR vs.\ GPT) and GPT's correctness.}
 \label{tab:correct}
 \vspace{.5em}
 \begin{tabular}{@{\extracolsep{.75cm}}lcc@{\extracolsep{.5cm}}cc}
   
   \toprule
    \quad
   & \multicolumn{2}{c}{\ \ \# Steps} \quad
   
   & \multicolumn{1}{c}{\ \ \ Correctness$^*$} \quad
   
   & 1$^{st}$ Wrong Step  \quad \\
   \cline{2-3}
    \quad
   & VECSR \quad
   &  GPT \quad
   & GPT  \quad
   & by GPT  \quad \\
     \midrule
   \gotosleeplinkv       \gotosleeplinkg       Go To Sleep &             3  &     5 &  83.33\% ($\pm$0.39)          &          \\
   \browseinternetlinkv  \browseinternetlinkg  Browse Internet &         8  &    10 &  \textit{50.00\%} ($\pm$0.52) & Step 2   \\
   \washteethlinkv       \washteethlinkg       Wash Teeth &             10  &    16 &  \textit{16.67\%} ($\pm$0.39) & Step 15  \\
   \brushteethlinkv      \brushteethlinkg      Brush Teeth &            10  &    21 &  91.67\% ($\pm$0.29)          &          \\
   \vacuumlinkv          \vacuumlinkg          Vacuum &                  5  &    10 &  83.33\% ($\pm$0.39)          &          \\
   \changesheetslinkv    \changesheetslinkg    Change Sheets [...]&     30  &    12 &  91.67\% ($\pm$0.29)          &          \\
   \washdirtydisheslinkv \washdirtydisheslinkg Wash Dirty Dishes &      12  &    15 &  83.33\% ($\pm$0.39)          &          \\
   \feedmelinkv          \feedmelinkg          Feed Me &                 8  &    10 &  \textit{33.33\%} ($\pm$0.49) & Step 2   \\
   \breakfastlinkv       \breakfastlinkg       Breakfast &               7  &    16 &  \textit{25.00\%} ($\pm$0.45) & Step 2   \\
   \readlinkv            \readlinkg            Read &                    6  &    8  &  \textit{66.67\%} ($\pm$0.49) & Unclear  \\
   \midrule
   \textbf{\hfill Average}                          &                   9.9 &  12.3 &  65.91\%          &          \\
   \bottomrule
 \end{tabular}
 \vspace{-.5em}
 \begin{center}\relsize{-0.5}
   ($^*$) Omitted VECSR's correctness because (by design) it is
   100\%. GPT's correctness is the mean score given by 12
   participants (in parenthesis, standard deviation).
 \end{center}
\end{table}

Huang et al. \cite{huang2022} evaluated the human-created action plans and found they were correct 70.05\% of the time. From our evaluation of the generated action plans, tasks rated close to or above 70\% correct are fairly accurate (for the task ``Read'', which was scored correct 66.67\% of the time, the task was correct enough that we were unable to identify the error). For the GPT-4o created plans we generated that scored lower than 70\% correct, we performed an analysis on where the plans stopped being correct. 
For example, the task which was rated the least correct was ``Wash Teeth'', which produced the following plan:
\begin{multicols}{2}
\relsize{-0.5}
\begin{enumerate}
    \item Walk to bathroom
    \item Walk to sink
    \item Find faucet
    \item Switch on faucet
    \item Grab toothbrush
    \item Find toothpaste
    \item Open toothpaste cap
    \item Squeeze toothpaste onto toothbrush
    \item Close toothpaste cap
    \item Put toothbrush into mouth
    \item Brush teeth thoroughly (front, back, sides, and chewing surfaces)
    \item Spit out toothpaste residue into sink
    \item Rinse toothbrush under faucet
    \item Switch off faucet
    \item Rinse mouth with water and spit
    \item Put toothbrush and toothpaste back in their place
\end{enumerate}
\end{multicols}

This action plan became incorrect in step 15 because the faucet is turned off by the time the action ``Rinse mouth with water and spit'' is suggested. Of interest, ``Wash Teeth'' and ``Brush Teeth'' are very similar tasks for which we used the same zero-shot prompt (substituting only ``Brush'' for ``Wash'' in front of ``Teeth''). However, the action plan generated for ``Wash Teeth'' was only 16.67\% correct, while ``Brush Teeth'' produced an action plan that was rated correct 91.67\% of the time. 
% New, shorter segment:
% For example, in the task ``Wash Teeth'' (rated 16.67\% correct) step 14 is ``Switch off faucet'' and step 15 is ``Rinse mouth with water and spit'', which is in the wrong order. Of interest, ``Wash Teeth'' and ``Brush Teeth'' are very similar tasks for which we used the same zero-shot prompt (substituting only ``Brush'' for ``Wash'' in front of ``Teeth''). However, unlike the former, the latter was rated correct 91.67\% of the time.

In most cases, the tasks generated by GPT-4o were longer and more verbose than those generated by VECSR. One notable exception is the ``Change Sheets and Pillow Cases'' task. This is the most complex task out of the ten examples, and the VECSR plan takes 18 more steps than the GPT-4o plan. This is due to the executability requirement of the VECSR system. For example, one of the instructions generated by GPT-4o is ``Take off the pillowcases from all pillows'', which would be multiple steps in the VirtualHome action format (``Walk to bedroom'' ``Walk to pillowcase 1'' ``Grab pillowcase 1'' ``Walk to pillowcase 2'' ``Grab pillowcase 2'').

Additionally, many of the other tasks incorrectly generated by GPT-4o contained a type of error that was commonly made in the human-created VirtualHome dataset, where an item is grabbed before it is ``found''. This susceptibility to flaws in the data is a hallmark of deep learning. By using our common sense reasoning system that connects directly to an embodied simulation environment, we see 100\% accurate results. Even though GPT-4o is a better LLM than the one used by Huang et al. \cite{huang2022}, it still leaves much to be desired in terms of correctness. And that is even ignoring the lack of context-awareness that make the LLM plans vague and unexecutable. With an LLM, even knowing that the data used for the zero-shot prompting will lead to incorrect answers, it is difficult or even impossible to correct this issue. The VECSR system is fully justifiable, as a proof of the reasoning behind the answer set for a query can be automatically produced (see Arias, Gupta, and Carro \cite{arias2021} for more information on the justification trees generated by s(CASP)). Thus, when errors are found in VECSR output, the system can easily be adjusted so that the error does not occur again. 

Going further, a noted weakness of the prior research is that while LLMs can provide steps for an action, they are not context-aware. The LLM has no access to the apartment when providing its steps, so it does not know if the items it is using to accomplish its goal even exist. In our testing with GPT-4o, it would often default to general types (``food'' instead of ``salmon'') or list possibilities (``breakfast food, such as cereal, bread, eggs, etc.''). These results are not necessarily incorrect, but it does limit the executability of those generated instructions. Even the more executable version proposed by Huang et al. \cite{huang2022} does not function with multiple objects of the same kind regardless of if more context was provided. Huang et al. evaluated a set of instructions as executable if the instructions used appropriate terminology (sit, grab <item>, etc.) rather than in larger chunks of more natural language (``Go over and sit on your comfortable chair'', ``Pick up the <item> if it is close by and available'', etc.). Although we compared VECSR output with those results, VECSR directly reasons with the environment, guaranteeing further that the objects being interacted with exist. 

Overall, VECSR output is more accurate and executable than that generated by an LLM. VECSR's output is also directly connected to the simulation environment.

\subsection{Generalizability}
\label{sec:generalizability}

Using s(CASP), VECSR generalizes from a small number of examples to perform well on unseen tasks. This is an improvement over neural networks, which require large quantities of data for decent performance on unseen data.
As mentioned above, VECSR was created considering only 10
tasks. This means that the common sense knowledge of all the actions
needed for correctly accomplishing these 10 tasks was added to its
knowledge base.
Once these actions are known, VECSR can perform any other task that
can be accomplished using a combination of the actions in its knowledge base.
To validate VECSR's ability to generalize from the learned tasks, we
randomly selected another 55 tasks from the VirtualHome database and asked VECSR to generate action sequences for them (given the knowledge base of actions corresponding to the original 10 tasks).
Of those 55 tasks, 39 tasks (70.91\%) were completed without
adding any additional rules or constraints to the knowledge
base. The results of this evaluation are available at \myurl{Examples/unseen_data_results.csv} on the repository. These preliminary
results reinforce our claim about the generalization of VECSR.

There are still gaps in the knowledge base due to actions that are
entirely unseen and cannot be generalized. For example, in six of the 16 uncompleted tasks (the 37.5\%) VECSR failed because none of the initial 10 tasks require opening something. These tasks include ``open door'' (which explicitly requires opening the door) and ``store meat in
freezer” (where the freezer must be opened to store the meat). But these tasks (and any tasks that require opening something) would be solvable, without affecting run-time performance, by adding the action ``open'' to the knowledge
base. We provide the required additions at \myurl{Examples/open_rules.pl}.
Note that compared to the amount of data consumed for comparable
performance by a deep learning system, this generalization capability
is very valuable.

\subsection{Time of Execution}
\label{sec:time}

For any common sense reasoning system to be useful in a practical sense, it has to be able to think on a timescale similar to humans \cite{lenat2023}. We implemented several optimizations in VECSR, described in Section \ref{sec:static}, to reduce the amount of time it takes to generate action plans. We tested the system with no optimizations, with each optimization listed in Section \ref{sec:static} individually, and all optimization applied together. Additionally, we timed the generation of plans by GPT-4o. The bottom of Section \ref{sec:static} gives more details on the specific optimizations and the online availability of the programs run for these results. While the programs were originally dynamically generated via connection with the VirtualHome simulation environment, they have been statically preserved so the experiment in this section can be recreated exactly. The VECSR results presented in this paper are from a base model 2023 MacBook Pro using version 0.24.01.31 of s(CASP).

\begin{table}[t!]
 \centering
 \caption{Comparison of run-time execution (in seconds).}
 \label{tab:time}
 \vspace{.5em}
 \begin{tabular}{@{\extracolsep{\fill}}lrrrrr@{\extracolsep{2em}}r}
   \toprule
     & Standard  &
     Modular  &
     Dep.Graph  &
     Part.Ground  & \quad Fully Opt. &
     GPT-4o  \\
   \midrule
    \gotosleeplinkf Go To Sleep & 197.41&   2.37&  229.95&    1.67\quad\quad&   \textbf{0.54}&  1.24\\
    \browseinternetlinkf Browse Internet &         -&  10.45& -&  2.19\quad\quad&  \textbf{0.68}& 3.23\\
    \washteethlinkf Wash Teeth &    552.87&   2.29&  278.73&   2.30\quad\quad&  \textbf{0.78}&  1.51\\
    \brushteethlinkf Brush Teeth &   414.96&   2.27&  -&   2.75\quad\quad&   \textbf{0.79}&  3.91\\
    \vacuumlinkf Vacuum &         391.97&  3.88&  62.68&  1.27\quad\quad&  \textbf{0.56}& 2.16\\
    \changesheetslinkf Change Sheets [...]&         -&  -&  -&   52.03\quad\quad&  14.16&  \textbf{5.50}\\
    \washdirtydisheslinkf Wash Dirty Dishes &     -&  -&  -&  5.06\quad\quad&  4.51&  \textbf{3.41}\\
    \feedmelinkf Feed Me &         -& 201.92& -& 4.48\quad\quad& \textbf{3.01}&  4.25\\
    \breakfastlinkf Breakfast &         -&  -&  -&  1.67\quad\quad&  \textbf{1.47}&  3.32\\
    \readlinkf Read &     485.45&   30.44&  557.06&   2.46\quad\quad&   \textbf{0.86}&  2.78\\
   \midrule
   \textbf{\hfill Average} &         408.53&  36.23&  282.11&   7.59\quad\quad&  \textbf{2.74}&  3.13\\
   \bottomrule
 \end{tabular}
 \centerline{\small Note: `-' means timeout after 10 minutes. Average
   does not take these tasks into account.}
 
\end{table}

Table \ref{tab:time} shows the execution times per example task for unoptimized (Standard) programs, each different optimization, a combination of all three, and the time it took for GPT-4o to generate answers. Even though some simple tasks could be executed without being optimized, even the simplest task still took well over three minutes (``Go To Sleep'' at 197.41 seconds) and 50\% of the tasks took over ten minutes. The most effective single optimization is partial grounding, which brought the average time for completion down from an average of almost 7 minutes among tasks that were completed at all to 7.59 seconds. However, 7.59 seconds is still too long for the average user, and the most complex task took nearly a minute to calculate a plan for (``Change Sheets and Pillow Cases'' at 52.03 seconds). However, by using all three optimizations, 60\% of the tasks execute in under a second. Programs fully optimized by our VECSR system generate a plan 0.39 seconds faster on average than GPT-4o.

This comparison does not include training time, which is much longer for an LLM (several months) than for a logic program. For VECSR, the cost is incurred in developing the knowledge base. Also, an LLM can handle any task, whether done correctly or not, while VECSR can only handle those tasks for which it has acquired common sense knowledge (although with a higher level of accuracy). 

\section{Conclusion}

VECSR represents a practical solution for the use of logic programming in high-fidelity real-world simulations. VECSR generates executable and correct action plans to accomplish high-level tasks and then perform those action plans in an embodied simulation environment. Using s(CASP)'s powerful reasoning capabilities and our novel framework we are able to outperform models that use LLMs as their common sense knowledge repository. With only a few sample tasks, we've created a model that is fast, accurate, and generalizable as a proof of concept of the power of logic programming for reasoning autonomy in embodied environments. 

Future work for VECSR can follow two main lines of thought: improvements to the system and further research built upon the VECSR foundation. In the future, more complex reasoning could be integrated, such as using event calculus \cite{vavsivcek2024} or reasoning over truly incomplete state information. Additionally, while LLMs are brittle alone there is promising research in pairing them with a symbolic system for improved explainability and reliability \cite{zeng2024}; which could be used for a natural language-enabled domestic helper for embodied environments. The ability to pair powerful reasoning with a connection to a simulated environment for embodied agents is a promising and practical framework for future research.

\nocite{*}
\bibliographystyle{eptcs}
\bibliography{generic}
\end{document}